\documentclass[aps,prb,twocolumn,superscriptaddress,groupedaddress]{revtex4}
\usepackage{graphicx}
\usepackage{latexsym}
\usepackage{amssymb}
\usepackage{amsmath}
\usepackage{amsfonts}
\usepackage{bm}
\usepackage{multirow}
\usepackage{color}
\usepackage{wasysym}
\newcommand{\ii}{\mathrm{i}}
\newcommand{\dd}{\mathrm{d}}

\newcommand{\hPsi}{\hat{\Psi}}
\newcommand{\hphi}{\hat{\phi}}

\newcommand{\hE}{\hat{E}}

\newcommand{\hA}{\hat{A}}

\newcommand{\hn}{\hat{n}}
\newcommand{\htheta}{\hat{\theta}}

\renewcommand{\Re}{\mathop{\mathrm{Re}}}
\renewcommand{\Im}{\mathop{\mathrm{Im}}}

\newcommand{\U}{\mathrm{U}}

\newcommand{\beq}{\begin{equation}}
\newcommand{\eeq}{\end{equation}}
\newcommand{\beqn}{\begin{eqnarray}}
\newcommand{\eeqn}{\end{eqnarray}}

\DeclareMathAlphabet{\mathbbold}{U}{bbold}{m}{n}

\def\U{{\rm U}}

\newcommand{\dtriangle}{\bigtriangledown}
\newcommand{\mcal}{\mathcal}
\newcommand{\mfrak}{\mathfrak}
\usepackage{ulem}
\usepackage[colorlinks=true,citecolor=blue,linkcolor=blue]{hyperref}
\usepackage{physics}

\let\grad\relax

\newcommand{\grad}{\nabla}

\newcommand{\hx}{\hat{x}}
\newcommand{\hy}{\hat{y}}
\newcommand{\hz}{\hat{z}}

\hypersetup{
colorlinks = true,
linkcolor=magenta,
citecolor=[rgb]{0.15,0.65,0.13},
urlcolor = [rgb]{0.25, 0.41, 0.88}}

\begin{document}

\title{A Multicritical Point with Infinite Fractal Symmetries}

\author{Nayan Myerson-Jain, Kaixiang Su, and Cenke Xu}
\affiliation{Department of Physics, University of California,
Santa Barbara, CA 93106, USA}

\begin{abstract}

Recently a ``Pascal's triangle model" constructed with $\U(1)$ rotor degrees of freedom was introduced, and it was shown that ({\it i.}) this model possesses an infinite series of fractal symmetries; and ({\it ii.}) it is the parent model of a series of $Z_p$ fractal models each with its own distinct fractal symmetry. In this work we discuss a multi-critical point of the Pascal's triangle model that is analogous to the Rokhsar-Kivelson (RK) point of the better known quantum dimer model. We demonstrate that the expectation value of the characteristic operator of each fractal symmetry at this multi-critical point decays as a power-law of space, and this multi-critical point is shared by the family of descendent $Z_p$ fractal models. Afterwards, we generalize our discussion to a $(3+1)d$ model termed the ``Pascal's tetrahedron model" that has both planar and fractal subsystem symmetries. We also establish a connection between the Pascal's tetrahedron model and the $\U(1)$ Haah's code. 
\end{abstract}

\maketitle

\section{Introduction} 

In recent years we have seen a very fruitful exploration of generalizing the concept of ``symmetry" in physics. It was demonstrated that introducing the concept of higher form symmetries is very useful for studying gauge fields~\cite{formsym0,formsym1,formsym2,formsym3,formsym4,formsym5,formsym6,formsym7,formsym8,mcgreevyreview}; in particular the analysis of the 't Hooft anomaly of higher form symmetries can lead to very clear and nontrivial conclusion for the low energy physics of gauge fields with a topological term~\cite{formanomaly}. Another set of generalized symmetries have their conserved charges defined on a subset of the system~\cite{fractonreview1,fractonreview2}, and these ``subsystem symmetries" can be further categorized into type-I, when the symmetric charges are defined on regular subsystems such as lines and planes, and type-II, when the symmetric charges reside on a fractal subsystem with often non-integer Hausdorff dimensions. So far systems with type-I subsystem symmetries are better understood, as not only can one study the lattice models with type-I subsystem symmetries, for gapless states (or phases) with type-I subsystem symmetry, it is also possible to describe the states with low energy effective theories which are constructed only with the low energy subspace of the Hilbert space rather than all the microscopic degrees of freedom of the system. One classic example of such was constructed by Paramekanti, Balents, Fisher in Ref.~\onlinecite{PBF} (dubbed the XY-plaquette model). The authors constructed a $(2+1)d$ quantum rotor model with one dimensional conservation of rotor numbers, and the authors studied in great detail a gapless state of this model. A low energy effective theory constructed with low energy modes of the system was developed for this gapless state, and a lot of physics can be studied systematically with the low energy effective theory. It was demonstrated that standard methods such as renormalization group (RG) can still be applied to the gapless state of the XY-plaquette model despite its highly unusual symmetries. The XY-plaquette model along with its low energy effective theory was later generalized to various models in higher dimensions, and it has attracted much enthusiasm recently~\cite{fractonreal1,ma,you,ye,gromov,seibergfracton1,seibergfracton2,seibergfracton3,seibergfracton4}. 

By the same standard, systems with type-II subsystem symmetries (or fractal symmetries), are much less understood. Much success was accomplished through studying exactly soluble lattice models such as the Haah's code~\cite{Haahcode}, but a low energy effective theory which does not involve all the microscopic degrees of freedom is more challenging compared with systems with type-I subsystem symmetries. Of course, a low energy effective theory may not be necessary if one only studies gapped states of type-II models, as under renormalization group (RG) flow gapped states are expected to ``flow to" an exactly soluble lattice model as the fixed point of the entire gapped phase, although a proper procedure of RG has not been developed for systems with fractal symmetries. But recent numerics suggests that a continuous quantum phase transition may exist for models with fractal symmetries~\cite{frankyizhi} (we note that earlier numerics suggested the opposite~\cite{juan1}), this calls for an understanding of gapless states with fractal symmetries at the same level as regular systems. A low energy effective theory will go a long way towards understanding the universal physics of such gapless systems without relying too much on the lattice models. 

\begin{figure}
\begin{center}
\includegraphics[width=0.35\textwidth, height = 2.25in]{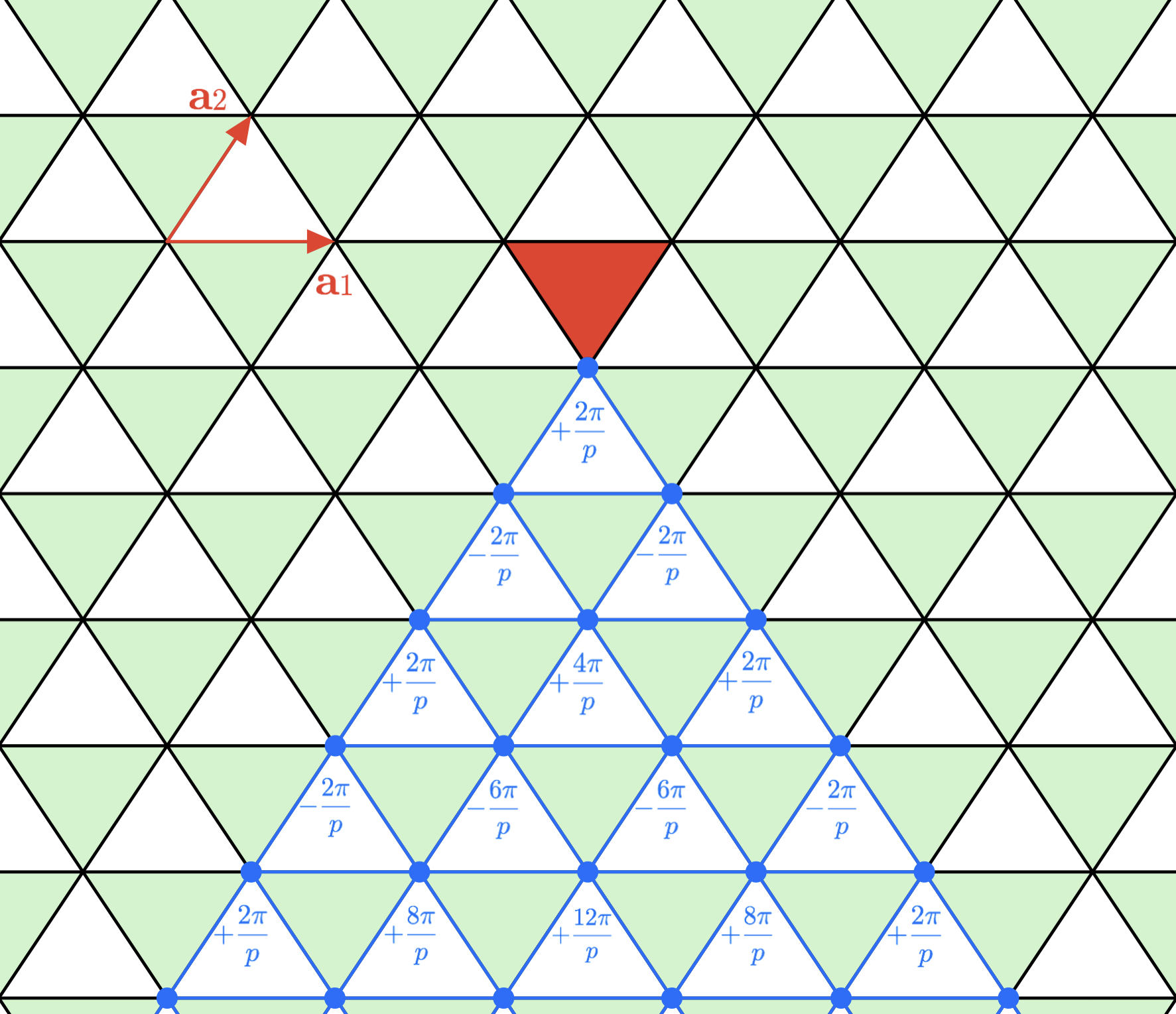}
\caption{The action of the Pascal's triangle symmetry on a patch (Eq.~\ref{Pascalpatch}) where the triangular region outlined in blue is $\mcal{R}(p)$. In the semi-classical version of the model, three fractonlike excitations localized at the triangular plaquettes at the corners of the Pascal's triangle are created (the top plaquette shown in red). The lattice sites are indexed by $(i_1,i_2)$ with lattice vectors $\vb{a}_1 = \hat{x}$ and $\vb{a}_2 = \hat{x}/2 + \sqrt{3} \hat{y}/2$.
}\label{Pascalpatchsym}
\end{center}
\end{figure}

In this work we construct a gapless state with an infinite series of fractal symmetries. The model we will start with was called the ``Pascal's triangle model":
\beqn
H = -K \sum_\dtriangle \cos(\htheta_1 + \htheta_2 + \htheta_3) + \frac{U}{2} \sum_{i } \hn_i^2, \label{Pascal}
\eeqn
where the boson phase $\htheta_i$ and boson number $\hn_i$ are a pair of conjugate variables which obey the usual algebra $[\htheta_i, \hn_j] = \ii \delta_{ij}$. This Hamiltonian exhibits a conventional $\U(1) \times \U(1)$ global symmetry involving rotating $\htheta$ on any two of the three sublattices of the triangular lattice by an opposite angle. This model was introduced for the purpose of understanding the quantum phase transition of the Sierpinski triangle model~\cite{frankyizhi}, but it was pointed out in Ref.~\onlinecite{Pascal} that this model actually supports a much larger group of fractal symmetries. In fact, the model possesses an infinite collection of distinct $Z_p$ fractal symmetries (one for each prime integer $p$) which together are called the ``Pascal's triangle symmetry" that involve a staggered rotation of the boson phase with support on a Pascal's triangle mod $p$. The action of this symmetry on a local patch of the lattice is generated by the so-called ``patch symmetry operator" of the Pascal's triangle symmetry
\beqn
U_p &=& e^{2 \pi \ii Q_P/p} \cr\cr
Q_p &=& \sum_{i \in \mcal{R}(p)} (-1)^{-i_2} { -i_2 \choose i_1} \hn_i \label{Pascalpatch}.
\eeqn
%\noindent
The sum in the patch charge $Q_p$ is over a triangular region $\mcal{R}(p)$ of length $L = p^{k}-1$ (centered at the origin, $\mcal{R}(p) \equiv \{ (i_1,i_2) : -L \leq i_2 \leq 0, 0 \leq i_1 \leq -i_2 \})$. The action of the patch symmetry operator is shown in Fig.~\ref{Pascalpatchsym}. It was demonstrated in Ref.~\onlinecite{Pascal} that for system sizes of $L = p^{k} - 1$, the $Z_p$ fractal symmetry becomes exact. The Pascal's triangle model also reduces to a family of descendent $Z_p$-clock model Hamiltonians that exhibit only a single one of its fractal symmetries. For example if one reduces the $\U(1)$ rotor degree of freedom to an Ising spin, the Pascal's triangle model reduces to the Sierpinski triangle model~\cite{NM,yoshida}.

\section{Multi-critical point of the Pascal's triangle model}

\subsection{Dual of the Pascal's triangle model}

The dual variables were introduced to understand the physics of the Pascal's triangle model~\cite{Pascal} Eq.~\ref{Pascal}: \beqn \sum_{j \in \bigtriangledown} \htheta_j = \hphi_{\bar{j}}, \ \ \ \ \hn_ j =- \sum_{\bar{j} \in \bar{\triangle} \ \mathrm{around} \ j}  \hPsi_{\bar{j}}. \label{eq:lattice_duality1}\eeqn
where $j, \bar{j}$ label the sites of the original and dual lattices respectively. $\hPsi$ is a discrete variable and $\hphi$ is its conjugate compact variable that satisfies the bosonic algebra $[\hPsi_{\bar{i}}, \hphi_{\bar{j}}] = \ii \delta_{ij}$. The Hamiltonian Eq.~\ref{Pascal} is now dual to a height model:
\begin{equation}
H_h = -K \sum_{\bar{j}} \cos(\hphi_{\bar{j}}) + \frac{U}{2} \sum_{\bar{\triangle}} (\hPsi_1 + \hPsi_2 + \hPsi_3)^2 \label{dual1}.
\end{equation}
We analyze this dual height model following the standard procedure. We first expand the height model to the quadratic order and add a vertex operator that re-enforces the compactness of $\hphi$ (or discreteness of $\hPsi$), then $\hPsi$ can be treated as a continuous variable: \beqn H_h^g \sim H_h - \alpha \sum_{\bar{j}} \cos(2 \pi \hPsi_{\bar{j}}). \eeqn A Euclidean action describing this model can be attained by expanding the Gaussian part of $H_h^g$ about the minima of the spectrum which occur at the $\pm \vb{K} = \pm (\frac{4 \pi}{3},0)$ points in the Brillouin zone (BZ). After expanding the height operator 
\beqn \Psi(\vb{r}) \sim e^{i \vb{K} \cdot \vb{r}} \psi(\vb{r}) + e^{-i \vb{K} \cdot \vb{r}} \psi^\ast(\vb{r}) \eeqn 
and taking the continuum limit, the low energy physics of the height model is supposedly described by the Lagrangian
\begin{equation}
\mcal{L} =  \frac{1}{2}( \partial_\tau \vec{\psi})^2 + \frac{\rho_2}{2} (\grad \vec{\psi})^2 - \alpha  \sum_{a} \cos( \vec{e}_a \cdot \vec{\psi}) \label{laggrad}. 
\end{equation}
To avoid potential confusion, we clarify that throughout this paper, the phrase continuum limit simply means the limit where the system size is taken to infinity while keeping a constant lattice spacing.
Eq.~\ref{laggrad} becomes an effective description of the lattice model when the field $\psi$ is a smooth function on the lattice. 
The two component vector field $\vec{\psi}$ is $\vec{\psi} = (\Re{\psi}, \Im{\psi})$ and $a = A, B, C$ label the three sublattices of the dual triangular lattice with $e_A = 2\pi (1, 0)$, $e_B = 2\pi(-1/2, \sqrt{3}/2)$, $e_C = 2\pi(-1/2, -\sqrt{3}/2)$. Analogous to the analysis of instantons in compact $\text{QED}_3$, due to the linear dispersion modes of the height field, the coefficient of the vertex operators $\alpha$ is strongly relevant. As a consequence, we conclude that the gapless modes of the Gaussian theory of the height model $H^g_h$ are unstable; or in other words the generic Pascal's triangle model should in general be in a gapped quantum disordered phase, just like the compact QED$_3$ should be in a confined phase without fine-tuning. The connection between the Pascal's triangle model and QED$_3$ will be made more explicit later. 

\subsection{Multi-critical point}

Although the vertex operators (the $\alpha$-terms) in Eq.~\ref{laggrad} destroy the gapless modes of the Gaussian theory of the dual height model, they can be suppressed to restore the gaplessness of the spectrum by fine-tuning the model to a multi-critical point. This multi-critical point has a simple description in terms of the dual height field:
\begin{equation}
\mcal{L} = \frac{1}{2} (\partial_\tau \vec{\psi})^2 + \frac{\rho_4}{2} (\grad^2 \vec{\psi})^2 - \alpha \sum_a \cos (\vec{e}_a \cdot \vec{\psi}) \label{finetunedlag}.
\end{equation}
This Lagrangian is two copies of the $z = 2$ quantum Lifshitz theory, which is pertinent to the study of the RK point of the quantum dimer model in $(2+1)d$ \cite{RK,dimerqed,dimer2,dimer3,dimer4}. 
This theory is the $(2+1)d$ analogue of the massless boson CFT in $(1+1)d$ and is known to be stable against vertex operators for some range of $\rho_4$. In particular, the $\beta$-function for the vertex operators to the leading order is
\begin{equation}
\beta(\alpha) = \bigg (4 - \frac{\pi}{2 \sqrt{\rho_4}} \bigg ) \alpha.
\end{equation}
\noindent
Later we will further demonstrate that this multi-critical point is shared by all the descendent $Z_p$ fractal models of the Pascal's triangle model. We can conclude that the vertex operators are irrelevant when $\rho_4 < \pi^2/64$ \cite{dimer2}.

Returning to lattice model Eq.~\ref{Pascal}, the multi-critical point can be reached by turning on nearest neighbor density-density interactions \beqn && \frac{J}{2} \sum_{\langle i,j \rangle} \hn_i \hn_j \sim \cr\cr && \frac{J}{2} \sum_{\langle \bar{\triangle}, \bar{\triangle}' \rangle}  (\hPsi_1 + \hPsi_2 + \hPsi_3)(\hPsi_{1'} + \hPsi_{2'} + \hPsi_{3'}) \eeqn which changes the band structure of the dual height model. A band structure calculation shows that by fine-tuning $J$ to certain point, the dispersion of the height field becomes quadratic. 

However, we must be careful since there are possible relevant operators and other marginal operators that may appear in RG that must be tuned to zero. These operators must obey the point-group symmetries of the lattice Hamiltonian. The scaling dimensions of space-time coordinates, and operators in the Lifshitz theory are $[\partial_x] = [\partial_y] = 1$, $[\partial_\tau] = 2$ and $[\psi_1] = [\psi_2] = 0$. An example of a relevant operator that must be tuned to zero is
\beqn
\mcal{L}_R = \alpha' \partial_y \psi \bigg ( \frac{\sqrt{3}}{2} \partial_x \psi + \frac{1}{2} \partial_y \psi \bigg ) \bigg ( \frac{\sqrt{3}}{2} \partial_x \psi - \frac{1}{2} \partial_y \psi \bigg ) + \text{c.c.} 
\label{relevant} \eeqn
This operator is similar to the one discussed in the context of quantum dimer model on the honeycomb lattice~\cite{dimer2}. Several other operators such as $\alpha'' (\grad \psi)^2 (\grad \psi^\ast)^2$ are marginal at the Lifshitz theory by power-counting, but they may become marginally relevant after considering quantum fluctuations~\cite{dimer4}. These perturbations had better also be tuned to zero to ensure the system stays at the multi-critical point. 
        
\subsection{Characteristic features at the multi-critical point}

To probe the physics at the multi-critical point, we would like to compute the expectation value of some characteristic operators. In particular we need an operator whose expectation value can distinguish different phases of a quantum many-body system. For a system with a symmetry, the so called ``patch operator" defined as symmetry transformation operation localized on a patch can serve as a diagnosis of the phase~\cite{jiwen}. This operator was also referred to as the ``disorder operator" for the example of Ising model in previous literature~\cite{kadanoff}, and later generalized to systems with type-I subsystem symmetries~\cite{XiaoChuanCatsym} as the order diagnosis operator. The application of the patch operator can be exemplified in the $1d$ quantum Ising model: the patch operator has short range, power-law, and long range expectation value in the Ising ordered phase, critical point, and the Ising disordered phase respectively (the exact opposite from the correlation function of the Ising order parameter). For more general cases, the expectation value of the patch operator should have a qualitative enhancement in a quantum disordered phase compared with the critical point. In the last few years the patch operators have been evaluated for various systems, it was shown that useful information can be extracted from the expectation value of the patch operator~\cite{meng1,xiaochuancat2,meng2,meng3}. 

Here we compute the expectation value of the patch operator of the Pascal's triangle symmetry Eq.~\ref{Pascalpatch}. As explained, opposed to the more standard order parameter, a long-range expectation value signifies a disordered phase. Power-law decay, however, still identifies a critical point. In terms of the dual variables of the height model, the Pascal's triangle symmetry charge on a patch $\mcal{R}(p)$ of length $r = p^{k}$ with the \textit{left corner} at the origin becomes
\beqn
Q_p(r) = \hPsi_{0,p^k} + (-1)^{p^{k}-1}\sum_{\bar{i}_1 = 0 }^{p^k} { p^{k} \choose \bar{i}_1 } \hPsi_{\bar{i}_1, 0}
\eeqn
where the support of the charge on a patch completely cancels in the bulk of the patch in the dual variables, and the remaining support is on the top corner and bottom row of the triangular region. {\it However, since ${ p^{k} \choose \bar{i}_1 }$ is always an integer multiple of $p$ for $\bar{i}_1 \neq 0, p^{k}$, the patch symmetry operator simplifies even further to involve only three $\hat{\Psi}$ operators at the corners of the patch:} 
\beqn
U_p(r) = \exp\left( \frac{2 \pi \ii}{p}(\hPsi_{0,p^k} + (-1)^{p^k-1} ( \hPsi_{0,0} + \hPsi_{p^k,0})) \right). \label{Pascalpatchdual}
\eeqn
Let us use lattice vectors of the triangular lattice $\vb{a}_1 = \hat{x}$ and $\vb{a}_2 = \hat{x}/2 + \sqrt{3} \hat{y}/2$ where $\hat{x}$ and $\hat{y}$ are the usual Cartesian unit vectors. $\hPsi_{0,r = p^k }$ is at the top corner of the triangle in the dual representation, while $\hPsi_{0,0}$ and $\hPsi_{r = p^k,0}$ are located at the bottom left and right corners of the Pascal's triangle respectively. For simplicity, we first consider the case where $p > 3$ when $(-1)^{r-1} = +1$. We will return to the case of $p = 2$ and $p = 3$ later. After expanding around the $\pm \vb{K} = \pm (4 \pi/3, 0)$ points we have that
\begin{align*}
    \hPsi_{0,0} &= \hPsi( \vb{r}_2 = 0) \sim  2 \psi_1(\vb{r}_2);\\
    \hPsi_{0,r} &= \hPsi(\vb{r}_1 = p^k \vb{a}_2)\\
    &\sim 2 \cos ( \frac{2 \pi p^k}{3} ) \psi_1(\vb{r}_1)  - 2 \sin ( \frac{2 \pi p^k}{3} ) \psi_2(\vb{r}_1);\\
    \hPsi_{r,0}  &= \hPsi(\vb{r}_3 = p^k \vb{a}_1)\\
    &\sim 2 \cos ( \frac{4 \pi p^k}{3} ) \psi_1(\vb{r}_3)  - 2 \sin ( \frac{4 \pi p^k}{3} ) \psi_2(\vb{r}_3).
\end{align*}

For $p \neq 3$, $\cos ( 2 \pi p^k/3 ) = \cos ( 4 \pi p^k/3) = - 1/2$ and $\sin ( 4 \pi p^k/3) = - \sin ( 2 \pi p^k/3) = \mp \sqrt{3}/2$. So we have after expanding $\hPsi_{0,0} +  \hPsi_{r,0} + \hPsi_{0,r} \sim [2 \psi_1(\vb{r}_2) - \psi_1(\vb{r}_1) -  \psi_1(\vb{r}_3)] 
\mp  \sqrt{3} [ \psi_2(\vb{r}_1) - \psi_2(\vb{r}_3)]$. 
Since $\psi_1$ and $\psi_2$ are uncoupled in the Lagrangian, the expectation of $U_p(r)$ factorizes into the product of two simpler correlators
\beqn
&&  \langle e^{ \frac{2 \pi \ii}{p} [ (2 \psi_1(x_2) - \psi_1(x_1) -  \psi_1(x_3))
    \mp  \sqrt{3} ( \psi_2(x_1) -  \psi_2(x_3) ) ]} \rangle \cr\cr 
&\sim& e^{ \frac{4 \pi^2}{p^2}[2 \langle \psi_1(x_2) \psi_1(x_1) \rangle + 2 \langle \psi_1(x_2) \psi_1(x_3) \rangle -  \langle \psi_1(x_1) \psi_1(x_3) \rangle ] }\cr\cr
&\times& e^{\frac{12 \pi^2}{p^2} \langle \psi_2(x_1) \psi_2(x_3) \rangle }.
\eeqn
Here the Euclidean propagator $\Delta(x - x') = \langle \psi_i(x) \psi_i (x') \rangle$ is computed according to a single copy of the Lifshitz theory. It is known that the propagator is logarithmic,
\beqn
\Delta(x - x') = \int^\Lambda \frac{\dd[3]{q}}{(2 \pi)^3} \text{ } \frac{e^{\ii q \cdot (x - x')}}{\omega^2 + \rho_4 q^4} = - \frac{\ln ( \Lambda \abs{x - x'})}{4 \pi \sqrt{\rho_4}} \label{propagator}
\eeqn
$\Lambda$ is the momentum cutoff. Since the propagator only depends on the distance $\abs{x - x'}$, and our fields lie at the corners of an equilateral of length $r$, $\langle U_p(r) \rangle$ becomes
\beqn
 \langle U_p(r) \rangle \sim  \exp\left( \frac{24 \pi^2}{p^2} \Delta(r) \right) \sim r^{ - \frac{6 \pi}{ p^2 \sqrt{\rho_4}}}.
\eeqn
Hence, the patch symmetry operator decays as a power-law with exponent $\delta_p = \frac{6 \pi }{p^2 \sqrt{\rho_4}}$. 

For $p =2$, the calculation is exactly the same but we must be careful since $(-1)^{r-1} = -1$ now. However, since the original height fields $\hPsi$ are quantized, as the operator $U_p(r) = e^{\ii \pi (\hPsi_{0,r} - (\hPsi_{0,0} + \hPsi_{r,0}))} = e^{\ii \pi (\hPsi_{0,r} + \hPsi_{0,0} + \hPsi_{r,0})}$, the latter operator having a sensible algebriac decay computable in the Lifshitz theory with exponent $\delta_2 = \frac{3 \pi}{2 \sqrt{\rho_4}}$.

Now for $p = 3$, $\hPsi_{0,0},\hPsi_{r,0},\hPsi_{0,r}$ all lie on the same sublattice. As a consequence, after expansion we have
\begin{align*}
    \hPsi_{0,0} &= \hPsi(\vb{r}_2 = 0) \sim 2 \psi_1(\vb{r}_2);\\
    \hPsi_{0,r} &= \hPsi(\vb{r}_1 = 3^k \vb{a}_2)  \sim  2 \psi_1(\vb{r}_1);\\
    \hPsi_{r,0}  &= \hPsi(\vb{r}_3 = 3^k \vb{a}_1) \sim  2 \psi_1(\vb{r}_3).
\end{align*}
Again making use of the fact that the initial fields were quantized we can compute the correlation of $U_3(r) = e^{\frac{2 \pi \ii}{3} (\hPsi_{0,r} + \hPsi_{0,0} + \hPsi_{r,0})} =  e^{\frac{2 \pi \ii}{3}(\hPsi_{0,r} + \hPsi_{0,0} - 2\hPsi_{r,0})} $. The latter form of the operator is an observable which has sensible decay once computed in the Lifshitz theory. This is because the sum of coefficients in front of the fields vanishes in this form, a necessary requirement to find a non-trivial result for the correlation function. After expansion, we must compute in the Lifshitz theories
\beqn
    && \langle e^{\frac{4 \pi \ii}{3} [\psi_1(x_1) + \psi_1(x_2) - 2\psi_1(x_3)]} \rangle \cr\cr 
    & \sim& e^{  \frac{16 \pi^2}{9} [2 \langle \psi_1(x_1) \psi_1(x_3) \rangle + 2 \langle \psi_1(x_2) \psi_1(x_3) \rangle - \langle \psi_1(x_2) \psi_1(x_1) \rangle]}, \cr\cr
    &=& e^{\frac{48 \pi^2}{9} \Delta(r)} \sim r^{ - \frac{4 \pi} {3 \sqrt{\rho_4}}}.
\eeqn
This is a slightly stronger decay than we found for $p \neq 3$, though the decay remains power-law. Hence, we conclude that the decay of the patch symmetry operator at the multi-critical point is $\langle U_p(r) \rangle \sim r^{-\frac{6 \pi}{p^2 \sqrt{\rho_4}}}$ with this exponent being twice as large for $p = 3$.

We can compare the behavior of the patch operator at the multi-critical point versus in the disordered phase, which as we argued before is the generic phase of the model Eq.~\ref{Pascal}. Inside the disordered phase, the vertex operator $\alpha$ is relevant, and $\psi$ is pinned at the minima of the vertex operators, with suppressed fluctuation. In this case, $\Delta(x - x') $ decays exponentially with distance, hence $\langle U_p(r) \rangle$ is expected to saturate to a finite value with long distance. This is consistent with the intuition one gather from other models: in the disordered phase, the disorder operator usually has enhanced correlation compared with the critical point.

\subsection{Multi-criticality of descendent $Z_p$ models}

We have shown that upon fine-tuning the Pascal's triangle model, it can reach a multi-critical point. We can consider if there exists a similar multi-critical point for the descendent fractal models with one of the fractal symmetries. For example, a descendent fractal model can be constructed by adding an operator to the theory that energetically favors the eigenvalues of the boson phase to be $\frac{2 \pi}{p} Z_p$ valued. The appropriately perturbed Pascal's triangle model is
\beqn
H_p = H -  g \sum_i \cos(p \htheta_i). \label{anisotropy}
\eeqn
Where $H$ is given by Eq.~\ref{Pascal}. If the $Z_p$ anisotropy is irrelevant at the multi-critical point discussed above, then we can conclude that all $Z_p$ fractal models attained by (weakly) breaking the rotor degrees of freedom down share the same multi-critical point as the Pascal's triangle model upon some fine-tuning. 

To analyze the relevance of the $Z_p$ anisotropy, we need to evaluate the perturbation of the $g$ operator in Eq.~\ref{anisotropy} at the multi-critical point discussed above. If an operator is relevant, it would make infrared divergent contribution to the partition function at certain order of perturbation. For example, in the XY-plaquette model, If one turns on the same $Z_p$ anisotropy $\sum_i g \cos(p \htheta_i)$, the fourth order perturbation of $g$ will have nonzero expectation value at the fixed point with $g = 0$, as long as the four operators are located at the four corners of a rectangle~\cite{PBF}, as required by the type-I subsystem symmetry. Though this expectation value decays with the size of the rectangle, it was shown that it can lead to infrared divergent contribution to the partition function, especially when the $Z_p$ operators are paired up as dipoles. Hence the $Z_p$ anisotropy can generate relevant dipole operators. But in our case, one can verify that, in order to preserve the fractal symmetries at the multi-critical point, at any finite order of perturbation of $g$, the expectation value $ \langle \cos ( p \htheta_{i_1}) \cdots\cos(p \htheta_{i_n}) \rangle $ always vanishes at large distance due to violation of the Pascal's triangle symmetry. Another way to look at the difference between our case and the XY-plaquette model is that, the nonzero four point correlation function of the XY-plaquette model can be viewed as the product of the $\cos(\nabla_x\nabla_y\htheta)$ term in the Hamiltonian over a rectangular area; but in our case a product of $\cos(\htheta_1 + \htheta_2 + \htheta_3)$ over certain area $A$ does not reduce to a simple finite composite operator when $A$ is large.

We stress here that, our argument for $g\cos(p \htheta)$ being irrelevant is based on perturbative expansion of $g$ at the fine-tuned fixed point where $\htheta$ takes continuous value. This argument is only supposed to hold for weak $Z_p$ anisotropy. 

\subsection{Relation of the Pascal's triangle model to a gauge theory}

Apart from the duality between the Pascal's triangle model and the height model, there exists another separate representation of the Pascal's triangle model in terms of a gauge theory. As we discussed in previous subsections, the dual height model of the Pascal's triangle model has two components of scalar fields at low energy, then following the standard ``boson-vortex" duality procedure~\cite{peskindual,halperindual,leedual}, the dual of the height model should be two $\U(1)$ gauge fields. In this section we explore the relation between the Pascal's triangle model and a $\U(1) \times \U(1)$ gauge theory. 

We define the canonical conjugate fields $\hat{E}_{i}$ and $\hat{A}_i$ on the sites of a honeycomb lattice as
\beqn && \hat{E}_{i \in \mcal{A}} = -(\hPsi_1+\hPsi_2+\hPsi_3)_{\bar{\triangle}}, \cr\cr 
&&\hat{E}_{i \in \mcal{B}} = -(\hPsi_1+\hPsi_2+\hPsi_3)_{\bar{\dtriangle}} \cr\cr
&& \hphi_{\bar{j}} = (\hat{A}_1+\hat{A}_2+\hat{A}_3+\hat{A}_4+\hat{A}_5+\hat{A}_6)_{\hexagon}
\label{eq:lattice_duality2}\eeqn
where $\mcal{A}$ and $\mcal{B}$ are the two sublattices of the honeycomb lattice. This honeycomb lattice has the original triangular lattice as one of its sublattices, say $\mcal{A}$. The electric fields $\hat{E}_i$ are forced to obey the Gauss' law constraint 
\beqn
G_{\hexagon} = \sum_{i \in \hexagon} \eta_i \hat{E}_i \equiv 0 \text{ with } \eta_i = 
\begin{cases}
+1 & i \in \mcal{A}\\
-1 & i \in \mcal{B}
\end{cases}.
\eeqn
The Hamiltonian of the resultant model, which we will refer to as the ``six-boson model", is
\beqn
H_{\text{sb}} = -K \sum_{\hexagon} \cos ( \sum_{i \in \hexagon} \hat{A}_i ) + \frac{U}{2} \sum_{i \in \mcal{A}} \hat{E}_i^2 + \sum_{i,i' \in \mcal{A}} \frac{J}{2} \hat{E}_{i} \hat{E}_{i'}  \label{six-boson}.
\eeqn

\begin{figure}
\begin{center}
\includegraphics[width=0.35\textwidth, height = 2.25in]{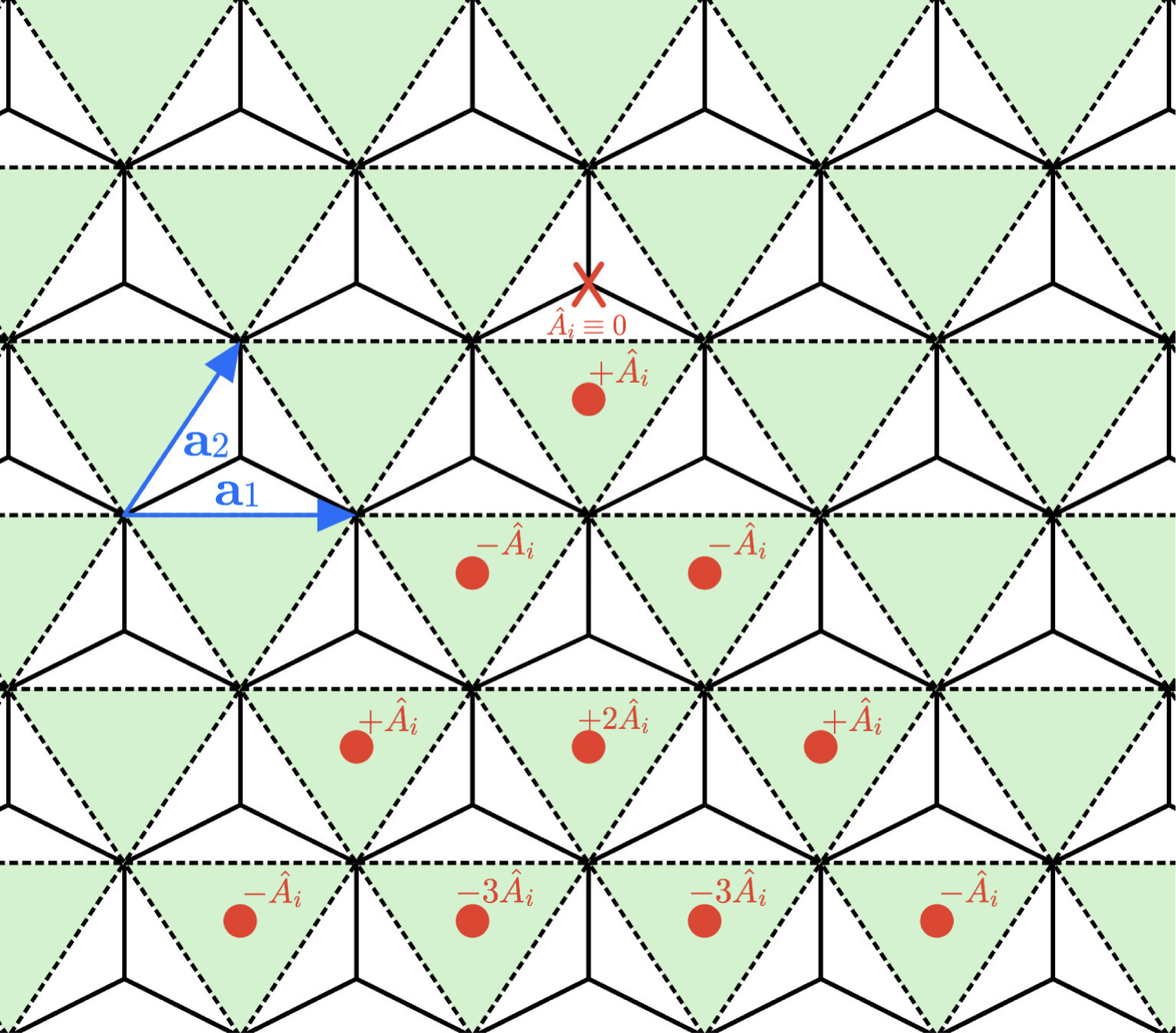}
\caption{This figure shows an example of a gauge transformation that fixes a single given $\hat{A}_{i \in \mcal{B}} \equiv 0$, and leaves the rest of the gauge fields on the $\mcal{B}$ sublattice intact. Successive gauge transformations of this form will fix all $\hA_{i \in \mcal{B}} = 0$. Such an accumulative gauge transformation that achieves this is given by Eq.~\ref{gaugechoice} for every $\alpha_{\hexagon}$ where $\tilde{\mcal{R}}(\infty)$ is centered at the top vertex of $\hexagon$.
}\label{gaugefixfig1}
\end{center}
\end{figure}

In addition, this six-boson model is directly related to the Pascal's triangle model with or without fine-tuning by gauge fixing. The gauge symmetry generated by the Gauss operator $G_{\hexagon}$ transforms the gauge fields according to
\beqn
\hat{A}_i \rightarrow \hat{A}_i + \eta_i \sum_{\hexagon \ni i} \alpha_{\hexagon} \label{gaugetransform}.
\eeqn
There is actually a valid gauge where the gauge fields on one of the two sublattices is completely gauged away. A gauge transformation that sets any single $\hat{A}_{i \in \mcal{B}}$ to zero and does not change any other $\hat{A}_{i \in \mcal{B}}$ is shown in Fig.~\ref{gaugefixfig1}. Using this gauge transformation, an accumulative gauge transformation can be constructed that fixes the gauge field to be zero at every site in the $\mcal{B}$ sublattice. This is achieved by making the following choice for every $\alpha_{\hexagon}$ appropriately:
\beqn
\alpha_{\hexagon} = \sum_{i \in \tilde{\mcal{R}}(\infty)} (-1)^{i_2} { i_2 \choose -i_1} \hat{A}_{i \in \mcal{B}}, \label{gaugechoice}
\eeqn
Specifically for $\alpha_{\hexagon}$ where $\hexagon$ is the honeycomb whose top corner is at the origin, the appropriate region being summed over is $\tilde{\mcal{R}}(\infty) = \{ (i_1,i_2) \in \mcal{B} : i_2 \geq \infty,  0 \leq -i_1 \leq  i_2 \}$.

Despite the form of the gauge transformation in Eq.~\ref{gaugetransform}, the gauge group of the six-boson model is actually $\U(1) \times \U(1)$. This should be expected from duality as each of the two scalar fields of the height model is dual to a $\U(1)$ gauge field in $(2+1)d$. As a consequence, there are two sets of Gauss' law constraints that generate two separate gauge transformations: one set where each $\hexagon$ lies on two of the sublattices of the dual triangular lattice and a second set where each $\hexagon$ lies on a different choice of two sublattices. The gauge group can also be directly verified in the continuum limit of the Gaussian approximation of Eq.~\ref{six-boson}. After the gauge fields are expanded around the $ \pm \vb{K}$ points in the BZ, the six-boson model reduces to two copies of QED$_3$ up to a field redefinition.

Other correlation functions of the Pascal's triangle model at the multi-critical point may be computed using the six-boson model. In particular, Wilson loops of the six-boson model end up reducing to correlation functions of products of the three-body operator $e^{i ( \htheta_1 + \htheta_2 + \htheta_3)}$ in the $\hA_{i \in \mcal{B}} \equiv 0$ gauge. These Wilson loops are given by staggered sums of the six-body interactions $\sum_{i \in \hexagon} \hA_i$ in some region so that the support in the bulk cancels and what remains is a loop of staggered gauge fields on the boundary. More precisely, 
\beqn
\langle W[\eta] \rangle_{\text{sb}} &=& \bigg \langle \exp ( i \sum_{\hexagon} \eta_{\hexagon} \bigg ( \sum_{i \in \hexagon} \hA_i \bigg ) )  \bigg \rangle_{\text{sb}} \cr\cr
&\overset{\hA_{i \in \mcal{B}} \equiv 0}{\longrightarrow}& \bigg \langle \prod_\dtriangle \exp ( i \eta_\dtriangle (\htheta_1 + \htheta_2 + \htheta_3) ) \bigg \rangle_{\text{Pascal}} \label{Wilson}
\eeqn
after a relabeling of $\hA_{i \in \mcal{A}}$ as $\htheta_i$. 

In addition, the $z = 2$ Lifshitz theory is known to have a duality to a gauge theory. In particular, the field theory of the fine-tuned height model at the multi-critical point is dual to
\beqn
\mcal{L} &=& \frac{1}{2 \rho_4} \bigg ( A_x^{(1)} \frac{\partial_\tau^2}{\partial_y^2} A_x^{(1)} + A_y^{(1)} \frac{\partial_\tau^2}{\partial_x^2} A_y^{(1)} \bigg ) + \frac{1}{2} (\nabla \times \vec{A}^{(1)})^2 \cr\cr
&+&(\vec{A}^{(1)} \leftrightarrow \vec{A}^{(2)}) \label{RKgauge}
\eeqn
for two $\U(1)$ gauge fields $\vec{A}^{(1)}$ and $\vec{A}^{(2)}$. This implies that the low energy theory of a fine-tuned version of the six-boson model, which reduces to the appropriate fine-tuned Pascal's triangle model, is equal to Eq.~\ref{RKgauge} up to a redefinition of the gauge fields. 

\begin{figure}
\begin{center}
\includegraphics[width=0.35\textwidth,, height = 2.25in]{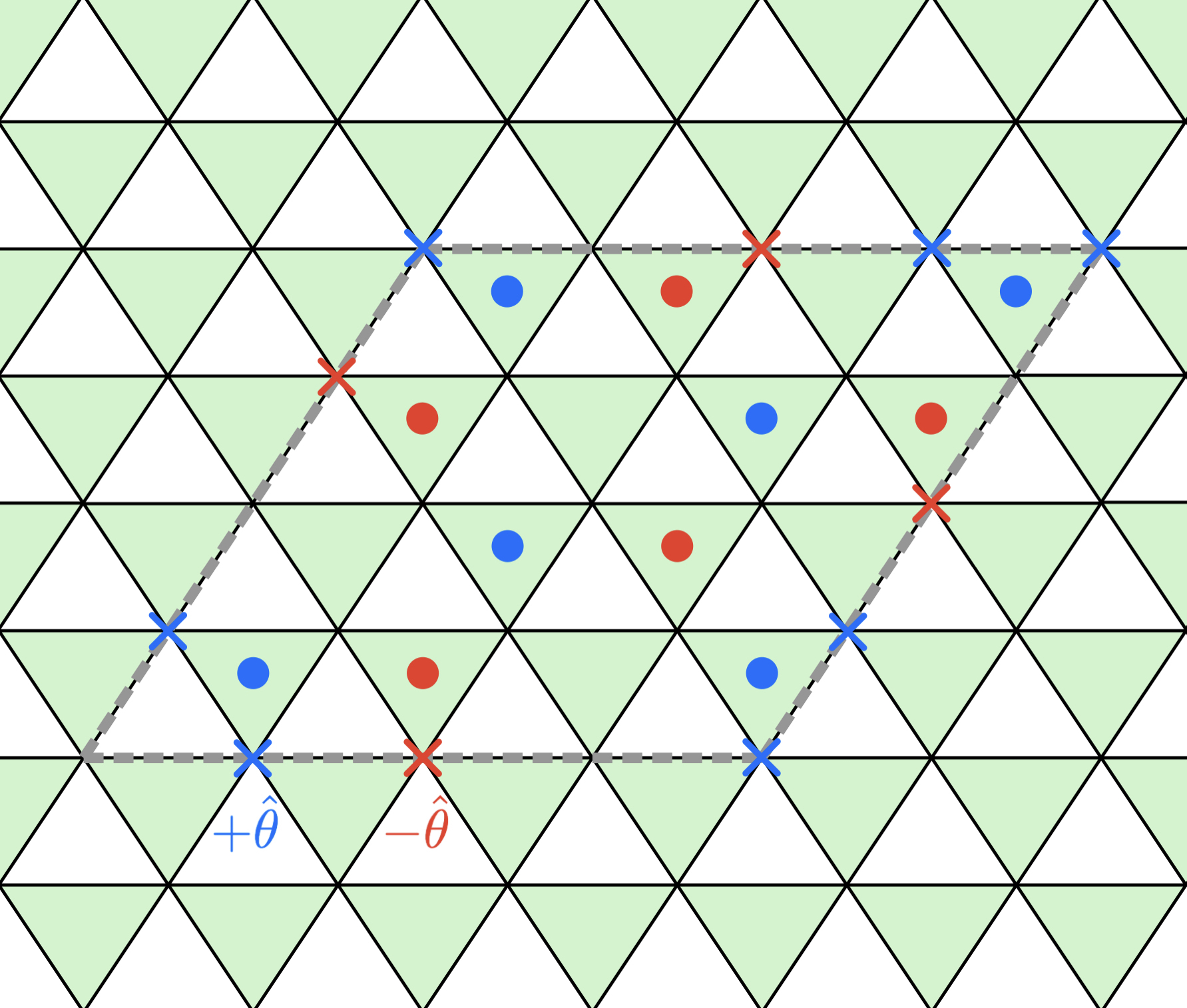}
\caption{ On the triangular lattice, a staggered product of $e^{i (\htheta_1+\htheta_2+\htheta_3)}$ and $e^{-i (\htheta_1+\htheta_2+\htheta_3)}$ on the $\tilde{A}$ and $\tilde{B}$ sublattices of the dual triangular lattice inside patch $\mcal{P}$ has support only on $\partial \mcal{P}$. The region inside the dashed gray line is $\mcal{P}$, with blue dots representing sites in $\tilde{A} \cap \mcal{P}$ and red dots representing sites in $\tilde{B} \cap \mcal{P}$. This product reduces to some staggered $\htheta$ variables leftover on $\partial \mcal{P}$ as shown by the blue and red crosses which is $e^{i \int_{\partial \mcal{P}} \gamma_i \htheta_i}$ in Eq.~\ref{Wilsonlooptriangular}.
}\label{Wilsonstaggerfig}
\end{center}
\end{figure}

It is known that the Wilson loops decay as a perimeter law in this theory, Eq.~\ref{RKgauge}. As this is the case, correlation functions of products of the three-body term in the Pascal's triangle model will also decay as a perimeter law. On the triangular lattice, we can consider a local patch let us denote as $\mcal{P}$. This patch contains a set of downward plaquettes which in themselves form the sites of the dual triangular lattice. If we now take a product of the three-body operators $e^{i(\htheta_1 + \htheta_2+ \htheta_3)}$ on two of the three dual sublattices of the plaquettes inside $\mcal{P}$ staggered on one of the sublattices, we recover an operator that is a product of $e^{\pm i  \htheta}$ with support only on $\partial \mcal{P}$. That is,

\beqn
\prod_{\dtriangle \in \tilde{A} \cap \mcal{P}} e^{i (\htheta_1+\htheta_2+\htheta_3)} \prod_{\dtriangle \in \tilde{B} \cap \mcal{P}} e^{-i (\htheta_1+\htheta_2+\htheta_3)} =  e^{ i \int_{\partial \mcal{P}} \gamma_i \htheta_i } \label{Wilsonlooptriangular}
\eeqn
where $\gamma_i = +1,-1,0$ depending on whether $\htheta_i$ on the boundary lies on a downward triangle on the $\tilde{A},\tilde{B},\tilde{C}$ sublattices of the dual triangular lattice. However, this operator is really just a Wilson loop of the six-boson model in the gauge $\hA_{i \in \mcal{B}} \equiv 0$ of the form Eq.~\ref{Wilson} and hence decays as a perimeter law with the number of boson phases on $\partial \mcal{P}$. This is explicitly shown in Fig. \ref{Wilsonstaggerfig}. But in the quantum disordered phase, which is the generic phase of either the Pascal's triangle model or the six-boson model, the Wilson loop will decay with an area law. In terms of the gauge theory language the quantum disordered phase is the confined phase. The area law can be shown by starting with the ground state with large $U$, and perform perturbation of $K$ operator in the Hamiltonian. 

\section{Pascal's tetrahedron model}

\subsection{The Hamiltonian and its symmetries}

There is a natural generalization of the Pascal's triangle model to $(3+1)d$ which we refer to as the Pascal's tetrahedron model. On the cubic lattice, the Hamiltonian of the Pascal's tetrahedron model reads
\beqn
H = -K \sum_i \cos(\sum_{\mu = 0, \hat{x}, \hat{y}, \hat{z}} \htheta_{i + \mu}) + \frac{U}{2} \sum_i \hn_i^2 \label{Pascal4}.
\eeqn
The four-body interaction $\cos(\htheta_i + \htheta_{i+\hx} + \htheta_{i+\hy} + \htheta_{i+\hz})$ involves the sum of boson phases over a tetrahedron on the cubic lattice; it is the natural generalization of the three-body interaction of Eq.~\ref{Pascal}. The ``Pascal's tetrahedron symmetries'' are also made up of an infinite series of $Z_p$ fractal symmetries. The action of the fractal symmetries on a finite tetrahedron with top corner at the origin of length $L = p^k-1$ is generated by the following patch symmetry operator

\beqn
U_p &=& e^{2 \pi \ii Q_p/p} \cr\cr
Q_p &=& \sum_{i \in \mcal{R}(p)} (-1)^{-i} \frac{(-i_x-i_y-i_z)!}{(-i_x)! (-i_y)! (-i_z)!} \hn_i. \label{Pascalpatchsym2}
%iz!/(iz-ix-iy)!ix!iy!
\eeqn
The action of this operator rotates the boson phase according to $\htheta_i \rightarrow \htheta_i + (2 \pi/p) (-1)^i (-i_x-i_y-i_z)!/(-i_x)!(-i_y)!(-i_z)!$ on the tetrahedral region $\mcal{R}(p) \equiv \{(i_x,i_y,i_z): i_x,i_y,i_z \leq 0, i_x+i_y+i_z \in [-L,0] \}$. Successive transformations of this form over extensively large regions will give rise to exact fractal symmetries, e.g. an exact fractal symmetry with $p = 2$ can be generated by three of these transformations for lattice size $L = 2^k-1$.

Despite Eq.~\ref{Pascal4} being the straightforward generalization of the Pascal's triangle model to $(3+1)d$, the Pascal's tetrahedron model manifests another type-I planar subsystem symmetry of which there is no analogue in $(2+1)d$. We may decompose the cubic lattice into four sublattices each spanned by vectors $\vb{b}_1 = \hat{x}+\hat{y}-\hat{z}, \vb{b}_2 = \hat{x}+\hat{z}-\hat{y}, \vb{b}_3 = \hat{y}+\hat{z}-\hat{x}$. The first sublattice is given by $\text{span}\{\vb{b}_1,\vb{b}_2,\vb{b}_3\}$ and the other three are obtained from subsequent shifts by $\hat{x},\hat{y}$ and $\hat{z}$ respectively. Now, consider a plane $\mfrak{L}$ on the cubic lattice with normal $\hat{n}_1 = \frac{1}{\sqrt{2}}(\hat{y}+\hat{z})$. We construct this plane by $\mfrak{L} = \mfrak{L}_1 \cup \mfrak{L}_2$ where $\mfrak{L}_1 = \text{span}\{\vb{b}_1+\vb{b}_2,\vb{b}_2\}$ and $\mfrak{L}_2$ is attained from $\mfrak{L}_1$ by a shift of $\hat{x}$. The Pascal's tetrahedron model is invariant under the following transformation on $\mfrak{L}$:
\beqn \U(1)^\text{(sub)} : \htheta_{i \in \mfrak{L}_1} \rightarrow \htheta_{i \in \mfrak{L}_1} + \alpha,  \ \ \theta_{i \in \mfrak{L}_2} \rightarrow \htheta_{i \in \mfrak{L}_2} - \alpha  \label{subU(1)} \eeqn 
Translations of the support of this transformation in the $\hat{y}$ and $\hat{z}$ directions will be exact symmetries as well. Furthermore, there exist two other sets of mutually orthogonal planes with normals $\hat{n}_2 = \frac{1}{\sqrt{2}} (\hat{x}+\hat{y}), \hat{n}_3 = \frac{1}{\sqrt{2}} (\hat{x}+\hat{z})$ of which a staggered rotation such as Eq.~\ref{subU(1)} leaves the Pascal's tetrahedron model invariant. These collection of planar transformations form the entire planar $\U(1)^{\text{(sub)}}$ symmetry of Eq.~\ref{Pascal4}. 

In the classical limit of the Pascal's tetrahedron model ($U = 0$), there exist both domain walls and fractonlike excitations. Firstly there exist fractonlike excitations created at the tetrahedra located at the corners of a Pascal's tetrahedron mod $p$ by acting with the fractal transformation in a local patch. Like the $(2+1)d$ case, these excitations cost $K[1 - \cos (2 \pi/p)]$ energy above the $\theta_i = 0$ configuration. In addition, there are one-dimensional loop excitations analogous to domain walls of the $(2+1)d$ $\text{XY}$ model created by acting with the $\U(1)^{(\text{sub})}$ symmetry on a patch of a plane such as $\mfrak{L}$. Starting with the uniform configuration $\theta_i \equiv 0$, these loop excitations cost an energy of $\sim N_\text{loop}K [ 1 - \cos \alpha]$ above such a ground-state configuration.

There is a dual of the Pascal's tetrahedron model similar to that of Eq.~\ref{dual1}. If we define variables on the dual cubic lattice whose lattice sites are located at the center of the cubes of the original lattice
\beqn
\htheta_i + \htheta_{i+\hx} + \htheta_{i+\hy} + \htheta_{i+\hz} &=& \hphi_{\bar{i}}, \cr \cr \hPsi_{\bar{i}} + \hPsi_{\bar{i}-\hx} + \hPsi_{\bar{i} - \hy} + \hPsi_{\bar{i}- \hz} &=& -\hn_i \label{dualmap41}.
\eeqn
Such a mapping preserves the bosonic algebra, and yields a dual height model of the Pascal's tetrahedron model
\beqn
H_h = -K \sum_{\bar{i}} \cos(\hphi_{\bar{i}}) + \frac{U}{2} \sum_{\bar{i}} \bigg ( \sum_{\mu = 0,\hx,\hy,\hz} \hPsi_{\bar{i}-\mu} \bigg )^2. \label{dual41}
\eeqn
This duality involving the $Z_2$, or Sierpinski-tetrahedron, version of this model was pointed out in Ref.~\onlinecite{juan2,YizhiCAutomata}. 
%Like the self-duality of the Pascal's triangle model, it involves the fractal symmetries as it maps the fractal symmetry generators to the identity. 

This dual height model does not explicitly possess the Pascal's tetrahedron symmetries of the original model as the $\hPsi$ fields are discrete, though it does retain a dual version of the $\U(1)^\text{(sub)}$ symmetry. This dual subsystem symmetry is realized in the spectrum via nodal lines of zero energy modes. It is straightforward to calculate the spectrum of the height model as $\epsilon(\vb{q}) = U[2+\cos(q_x) + \cos(q_y) + \cos(q_z) + \cos(q_x-q_y) + \cos(q_x - q_z) + \cos(q_y - q_z)]$. Despite the modest form of the Hamiltonian and the spectrum, we know that the height model has a dual $\U(1)^{(\text{sub})}$ symmetry and so its spectrum must display nodal lines where $\epsilon(\vb{q}) = 0$. In fact there are six nodal line segments in the BZ, e.g. $\{\vb{q} = (t - \pi,t,\pi) : t \in [0,\pi] \}$. If expanded along this line, the dispersion of the height field becomes \beqn
\epsilon(\vb{q}) &\sim& -2U q_xq_y + 2 U\cos(t) (q_x q_z - q_y q_z) \cr\cr &+& U(q_x^2+q_y^2 + q_z^2).
\eeqn  As a consequence, an analysis of a potential multicritical point of the Pascal's tetrahedron model must involve a similar height field expansion as in the analysis in $(2+1)d$, but around each of the nodal lines. We leave the detailed discussion of the RK point of tetrahedron model to future studies. 

\begin{figure}
\begin{center}
\includegraphics[width=0.35\textwidth, height = 2.75in]{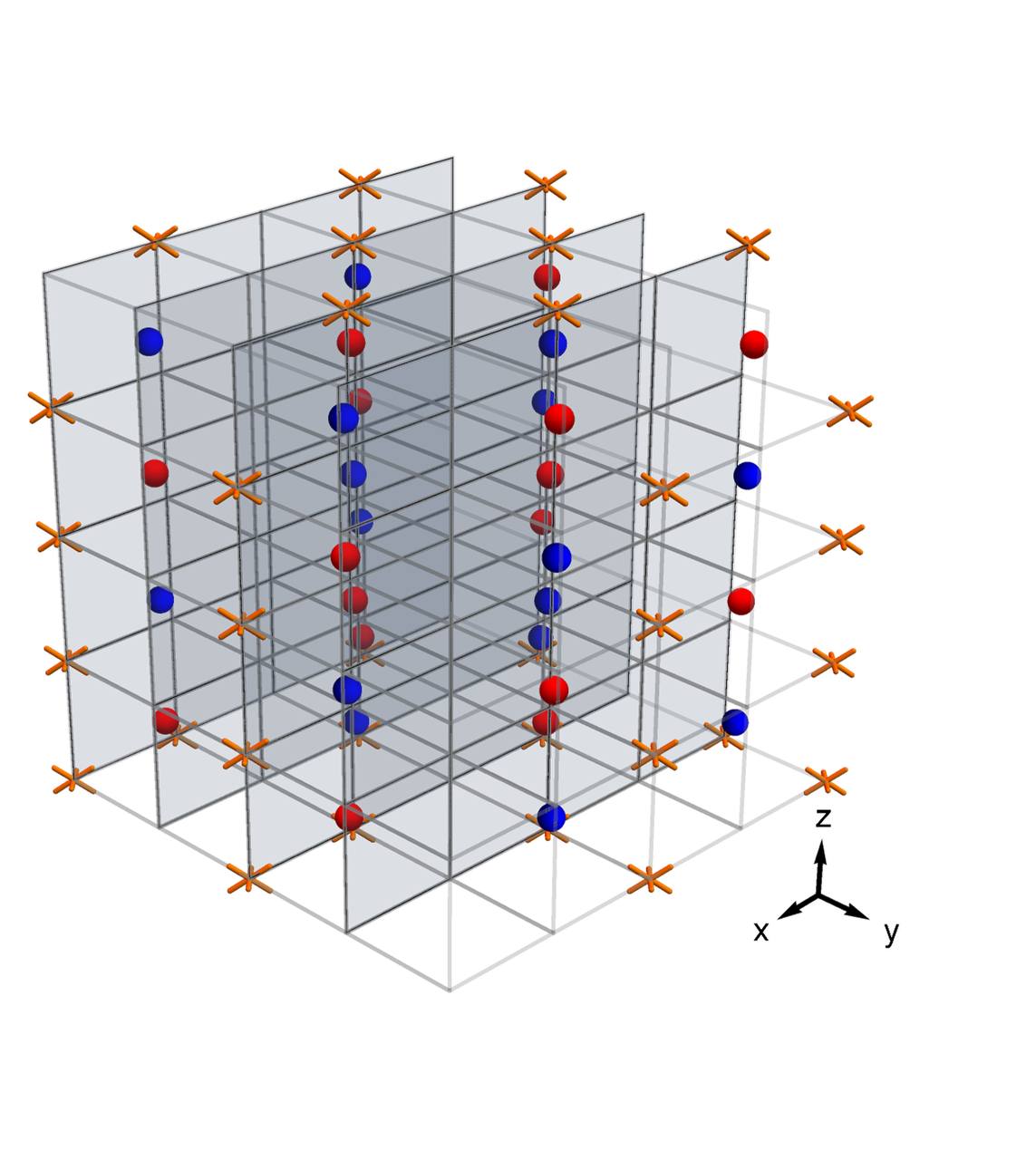}
\includegraphics[width=0.35\textwidth, height = 2.75in]{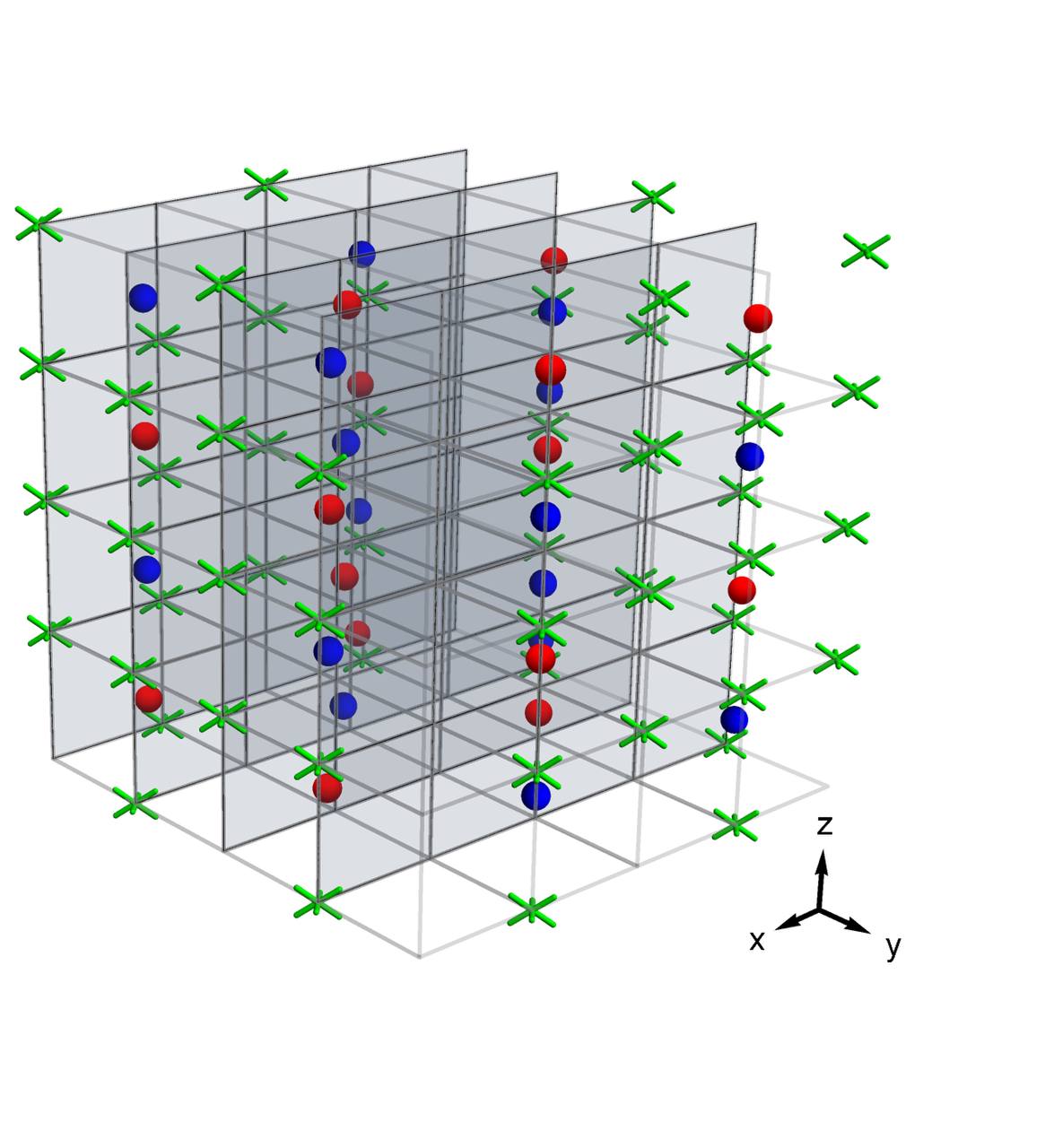}
\caption{By making a staggered product of $\exp( \ii ( \hA^{(1)}_i + \dots+ \hA^{(2)}_{i+\hy+\hz}))$ term in Eq.~\ref{dual43} on the cubes labelled as blue and red which belong to different dual sublattices, both gauge fields $\hA^{(1)}$ and $\hA^{(2)}$ cancel out in the bulk; the remaining fields are located on the boundary. The remaining $\hA^{(1)}$ and $\hA^{(2)}$ fields are labelled as crosses in the top and bottom figures respectively. }\label{Haahwilson}
\end{center}
\end{figure}

\subsection{Relation to the U(1) Haah's code}

Like the Pascal's triangle model, the Pascal's tetrahedron model is related to a gauge theory by gauge fixing. However, while in the $(2+1)d$ this theory was a conventional $\U(1) \times \U(1)$ gauge theory, we will show that the appropriate gauge theory in $(3+1)d$ is the pure (i.e. uncoupled from matter, the Gauss' operator is non-dynamical) version of the $\U(1)$ Haah's code. The dual height model in Eq.~\ref{dual41} has a second duality. We can define two new electric and gauge fields according to
\beqn
\hE_i^{(1)} &=& - (\hPsi_{\bar{i}} + \hPsi_{\bar{i}-\hx} + \hPsi_{\bar{i}-\hy} + \hPsi_{\bar{i}-\hz}), \cr\cr
\hE_i^{(2)} &=& - (\hPsi_{\bar{i}} + \hPsi_{\bar{i}-\hx-\hy} + \hPsi_{\bar{i}-\hx-\hz} + \hPsi_{\bar{i}-\hy-\hz}), \cr\cr
\hphi_{\bar{i}} &=& \hA^{(1)}_i + \hA^{(1)}_{i+\hx} + \hA^{(1)}_{i+\hy} + \hA^{(1)}_{i+\hz}  \cr\cr &+& \hA^{(2)}_{i} + \hA^{(2)}_{i+\hx+\hy} + \hA^{(2)}_{i+\hx+\hz} + \hA^{(2)}_{i+\hy+\hz}.
\eeqn
The $Z_2$ version of this mapping is equivalent to the ``F-S'' duality outlined in Ref.~\onlinecite{Sagarduality}. There exists a linear combination of interactions of the form $(\hPsi_{\bar{i}}  + \hPsi_{\bar{i}-\hx} + \hPsi_{\bar{i}-\hy} + \hPsi_{\bar{i}-\hz})$ and $(\hPsi_{\bar{i}}  + \hPsi_{\bar{i}-\hx-\hy} + \hPsi_{\bar{i}-\hx-\hz} + \hPsi_{\bar{i}-\hy-\hz})$ that vanishes, and as a result the electric fields of the dual must obey the corresponding Gauss' law constraint this time being
\beqn
G_i &=& \hE^{(1)}_{i+\hx+\hy+\hz} + \hE^{(1)}_{i+\hx} + \hE^{(1)}_{i+\hy} + \hE^{(1)}_{i+\hz}  \cr\cr &-& \hE^{(2)}_{i+\hx+\hy+\hz} - \hE^{(2)}_{i+\hx+\hy} + \hE^{(2)}_{i+\hx+\hz} - \hE^{(2)}_{i+\hy+\hz} \equiv 0 .
\eeqn
The Hamiltonian of the resulting gauge theory takes the form of the rotor limit of pure Haah's code

\beqn
H &=& -K \sum_i \text{cos}(\hA^{(1)}_i + \hA^{(1)}_{i+\hx} + \hA^{(1)}_{i+\hy} + \hA^{(1)}_{i+\hz}  \cr\cr &+& \hA^{(2)}_{i} + \hA^{(2)}_{i+\hx+\hy} + \hA^{(2)}_{i+\hx+\hz} + \hA^{(2)}_{i+\hy+\hz}) \cr\cr
&+& \frac{U}{2} \sum_i (\hE_i^{(1)})^2  \label{dual43}.
\eeqn
Reducing the degrees of freedom to Ising spins via $(\sigma^z)^{(a)}_j = e^{i \hat{A}^{(a)}_j}$ and $(\sigma^x)^{(a)}_j = e^{\pi i \hat{E}_j^{(a)}}$, leaves us with a perturbed version of the pure $Z_2$ Haah's code. Fractional gauge fluxes of this model created at the corners of products of $e^{i \hE_i^{(1)}}$ or $e^{i \hE^{(2)}_i}$ in the shape of Pascal's tetrahedra mod $p$ will be completely immobile fractons. As each of these fractons costs an energy $\Delta E = K[1 - \cos (2 \pi/p)]$ in the classical limit ($U = 0$). 

The Gauss operator generates a gauge transformation acting on the gauge fields of
\beqn
\hA^{(1)}_i &\rightarrow& \hA^{(1)}_i + \alpha_{i-\hx} + \alpha_{i-\hy} + \alpha_{i-\hz} + \alpha_{i-\hx-\hy-\hz}, \cr\cr 
\hA^{(2)}_i &\rightarrow& \hA^{(2)}_i - \alpha_{i-\hx-\hy} - \alpha_{i-\hx-\hz} - \alpha_{i-\hy-\hz} - \alpha_{i-\hx-\hy-\hz}.  \label{gaugetransform2}
\eeqn
The pure $\U(1)$ Haah's code can be reduced to the Pascal's tetrahedron model upon gauge fixing and tuning of the electric couplings. We may choose
\beqn
\alpha_{j - \hx - \hy - \hz} = \sum_{i \in \tilde{\mcal{R}}(\infty)} (-1)^{i} \frac{(i_x+i_y+i_z)!}{i_x!i_y!i_z!} \hA_i^{(2)},
\eeqn
with tetrahedral region $\tilde{\mcal{R}}(\infty)$ being that of $\mcal{R}(\infty)$ with the opposite orientation centered at site $j$. Such a transformation sets $\hA^{(2)}_j \equiv 0$ at every site as with this choice, $\alpha_j+\alpha_{j+\hx} + \alpha_{j+\hy} + \alpha_{j+\hz} = \hA^{(2)}_j$ yielding exactly the Pascal's tetrahedron model. This should be expected given the dualities we have just described; our pure $\U(1)$ Haah's code is dual to the height model Eq.~\ref{dual41} which in turn is dual to Eq.~\ref{Pascal4}, and hence must be dynamically equivalent to the Pascal's tetrahedron model.

Since the height field in the dual representation of the tetrahedron model is a scalar field, then one would expect that the dual of a scalar field is a 2-form gauge field in three dimensional space. And a 2-form gauge field should have ``Wilson membranes" rather than Wilson lines. Indeed, one can show that gauge invariant operators constructed as staggered products of the $K$-operator, $\exp(\ii (\hA_{i}^{(1)} + \dots + \hA_{i+\hy+\hz}^{(2)}))$ in Eq.~\ref{dual43}, in a three dimensional bulk of the system reduces to surface products of $\hA^{(1)}$ and $\hA^{(2)}$ (Fig.~\ref{Haahwilson}). Specifically, one can consider a box shaped patch of the cubic lattice. As each cube inside the patch belongs to one of the four sublattices of the dual cubic lattice, a product of $K$-operators staggered on two of the dual sublattices reduces to an operator which only has remaining support on the boundary of the patch. This operator appropriately reduces to a product of boson phases in the boundary of the patch in the Pascal's tetrahedron model upon gauge fixing.

\section{Discussion}

In this work we constructed a gapless multi-critical point from a lattice model with an infinite series of fractal symmetries, which together are called the Pascal's triangle symmetry. This multi-critical point can be reached by tuning a finite number of parameters of the generic Pascal's triangle model, and it is analogous to the RK point of the quantum dimer model in the sense that it is gapless with a $z = 2$ dispersion. At this multi-critical point, physics can be analyzed conveniently in the dual description of the system; and we demonstrate that the characteristic ``patch operators" of all the fractal symmetries have an expectation value that decay with a power law of space, which is usually a sign of criticality. Potential generalizations to three dimensional systems are also discussed, but we leave an in-depth discussion of a potential RK point of the Pascal's tetrahedron model to a future work. As we mentioned earlier, gapless states with type-I subsystem symmetries have been studied quite extensively since Ref.~\onlinecite{PBF}, we hope our work can draw attention to the study of gapless systems with type-II fractal symmetries.  

We would like to acknowledge Sagar Vijay and Shang Liu for insightful discussions. This work is supported by NSF Grant No. DMR-1920434, and
the Simons Investigator program. Part of the discussions in this work originated from an unpublished work with Ashvin Vishwanath many years back. 

%\section{Acknowledgements}

\bibliography{fractal}

\appendix

\section{Review of the XY-plaquette Model}

When we discuss gapless states with subsystem symmetries, it is always useful to first think about the XY-plaquette model~\cite{PBF}. The XY-plaquette model is a quantum rotor model with a type-I subsystem symmetry. The Hamiltonian of the model on the square lattice reads
\beqn
H_{\text{plaquette}} = -K \sum_{\square} \cos( \nabla_x \nabla_y \htheta ) + \frac{U}{2} \sum_{i} \hn_i^2 \label{PBF}.
\eeqn
The operator $\nabla_x\nabla_y \htheta_i$ is defined as $\sum_{i \in \square}  (-1)^{i} \htheta_{i} $ on each unit square plaquette. This model admits a $\U(1)^{(\text{sub})}$ symmetry involving rotating the boson phase along lines of constant $i_x$ or $i_y$:
\beqn \U(1)^\text{(sub)} &:& \theta_{i_x,i_y = \text{const}} \rightarrow  \theta_{i_x,i_y = \text{const}}+ \alpha(i_y), \cr\cr
\text{ or } &:& \theta_{i_x = \text{const},i_y } \rightarrow  \theta_{i_x = \text{const},i_y} + \alpha(i_x).  \label{PBFU(1)} \eeqn 
As per usual, the boson number and phase obey the canonical commutation relation $[\htheta_i,\hn_j] = \ii \delta_{ij}$. This model is known to exhibit multiple dual representations. We will discuss two of these here as they are analogous to the dualities involving the Pascal's triangle model discussed in the main text. Firstly, there is a dual to the height model that can be attained by defining variables that live on the dual square lattice 
\beqn \sum_{j \in \square} (-1)^j \htheta_j = \hphi_{\bar{j}}, \ \ \ \ \hn_ j =- \sum_{\bar{j} \in \bar{\square}  \text{ around} \ j}  (-1)^{\bar{j}}\hPsi_{\bar{j}}. \eeqn
The $\bar{j}$ label the sites of the dual square lattice and live at the center of the plaquettes of the original square lattice. The dual Hamiltonian is a height version of the XY-plaquette model
\beqn
H_h = - K \sum_{\bar{j}} \cos(\hphi_{\bar{j}}) + \frac{U}{2} \sum_{\bar{\square}} \bigg ( \nabla_x\nabla_y \Psi \bigg )^2 \label{HeightPBF}
\eeqn
This height dual exhibits a dual $\U(1)^{(\text{sub})}$ symmetry -- shifting the height fields $\hPsi_{\bar{j}}$ by an integer along a line of constant $\bar{j}_x$ or $\bar{j}_y$ leaves the Hamiltonian invariant. 

The XY-plaquette model has a gapless phase that allows us to take a Gaussian expansion of the cosine term in Eq.~\ref{PBF}. The (continuum limit) effective theory of the Gaussian theory reads
\beqn
\mcal{L}_{\text{plaquette}} = \frac{1}{2} (\partial_\tau \theta)^2 + \frac{\rho}{2} (\partial_x \partial_y \theta)^2 \label{PBFLag}.
\eeqn
We remind the readers that in the current paper the continuum limit means infinite system size, constant lattice spacing. Due to the special subsystem symmetry of the XY-plaquette model (and its various generalizations), one {\it cannot} take lattice spacing $a$ to zero, which would lead to unphysical conclusions. This effective theory is manifestly non-Lorentzian as it retains part of the type-I subsystem symmetry: the Lagrangian is invariant under any shift \beqn \theta(x,y,\tau) \rightarrow \theta(x,y,\tau) + f_1(x) + f_2(y) , \eeqn where $f_1$ and $f_2$ are completely arbitrary functions. This Gaussian theory is gapless, as the dispersion of the modes is $\omega(\vb{q}) \sim \sqrt{\rho} \abs{q_x q_y}$. The subsystem symmetry of the effective theory does {\it not} necessarily include the entire subsystem symmetry of the original lattice model, as $f_1$ and $f_2$ are presumably smooth functions of space. In fact, some of the low energy modes of the original lattice near the boundary of the Brillouin zone are ignored in the effective theory. 

There is another dual of the height model Eq.~\ref{HeightPBF} that is relevant for our discussion. Defining new variables on the links of the \textit{original} square lattice
\beqn 
\hE_{j,j+\hx} &=& \nabla_y \hPsi_{\bar{j}},  \ \ \hE_{j,j+\hy} = - \nabla_x \hPsi_{\bar{j}}, \cr\cr  \hphi_{\bar{j}} &=& (\nabla \times \hA_{j,j+\mu})_{\text{around } \bar{j}}.
\eeqn
The electric fields $\hE$ are forced to obey the usual Gauss' law constraint of $\text{QED}_3$,
\beqn
G_j = (\nabla_x \hE_x + \nabla_y \hE_y)_j \equiv 0
\eeqn
The dual Hamiltonian, consequently is a fine-tuned version of the compact $\text{QED}_3$
\beqn
H = -K \sum_\square \cos(\nabla \times \hA)_\square + \frac{U}{8} \left(\nabla_x \hE_x - \nabla_y \hE_y \right)^2. \label{PBFgauge}
\eeqn
To bridge the connection between the two representations of the XY-plaquette model, i.e. Eq.~\ref{PBF} and Eq.~\ref{PBFgauge}, we need to fix the gauge of $\hA$. We choose a gauge such that $\nabla_y \hA_x + \nabla_x \hA_y = 0$, which can always be met by choosing a right gauge transformation $\hA \rightarrow \hA + \nabla \varphi$. For example, for a configuration of $\hA$, one can choose function $\varphi(x,y)$ to be $ - \frac{1}{2}\int_0^x \int_0^y dx' dy' (\nabla_x \hA_y + \nabla_y \hA_x)  $. Then the relation between Eq.~\ref{PBF} and Eq.~\ref{PBFgauge} is that $ \nabla_x \htheta = 2 \hA_x$, $\nabla_y \htheta = - 2 \hA_y$, $2 \hn = \nabla_x \hE_x - \nabla_y \hE_y$. Note that since $\nabla_x \hE_x = - \nabla_y \hE_y $ due to the Gauss law constraint, $\hat{n}$ takes its value in all integers. The subsystem transformation $\htheta(x) \rightarrow \htheta(x) + f_1(x) + f_2(y)$ becomes {\it part of} the residual gauge transformation: $ \hA \rightarrow \hA + \nabla \varphi$, where $\varphi = ( f_1(x) - f_2(y))/2$. In particular, the nonzero four point correlation function of $\cos(\ii \htheta)$ where $\htheta$ locate on four corners of a rectangle becomes a Wilson loop of the gauge field $\hA$.

\end{document}